\documentclass[aps,prl,twocolumn,superscriptaddress]{revtex4-2}
\usepackage{graphicx}
\usepackage{amsmath}
\usepackage{amssymb}
\usepackage{color}
\usepackage{physics}
\usepackage{hyperref}

\begin{document}

\title{Non-equilibrium coupling to a diffusing density breaks Ising universality}

\author{Mattia Scandolo}%
\affiliation{%
    Laboratoire de Physique de l’École normale supérieure, ENS, Université PSL, CNRS, Sorbonne Université, Université de Paris, F-75005 Paris, France
}%
\affiliation{%
    Department of Physics, University of Chicago, Chicago, Illinois 60637, USA
}%
\author{Johannes Pausch}
\affiliation{%
    Department of Mathematics, Imperial College London, London SW7 2AZ, United Kingdom
}%
\author{Mike E. Cates}
\affiliation{%
    DAMTP, Centre for Mathematical Sciences, University of Cambridge, Cambridge CB3 0WA, United Kingdom
}%
\author{Luca Di Carlo}%
\affiliation{%
    Joseph Henry Laboratories of Physics, Princeton University, Princeton, New Jersey 08544, USA
}
\affiliation{%
    Lewis-Sigler Institute for Integrative Genomics, Princeton University, Princeton, New Jersey 08540, USA
}

\date{\today}

\begin{abstract}
The Ising universality class is remarkably robust to non-equilibrium perturbations, which generically flow to zero under renormalization. We show that this robustness fails when an order parameter is coupled nonreciprocally to a conserved diffusive density. Below $d_c=4$, the renormalization group flows to a fast-diffusion fixed point at which the density acts as a long-range multiplicative noise, producing a novel universality class. The non-equilibrium nature of the fixed point is manifest in the large-scale violation of the fluctuation-dissipation relations, reflected in a splitting of the scaling exponents of the two-point correlation and response functions--a measurable hallmark of non-equilibrium critical fluctuations. A two-loop calculation establishes the stability of this fixed point but yields a small correction-to-scaling exponent $\omega\approx0.020$ in $d=3$, implying strong finite-size corrections. An all-orders modified Harris criterion $\nu>2/(d+z-2)$ confirms that the BIM fixed point governs criticality in $d=3$, with Ising universality recovered only at $d=2$.
\end{abstract}

\maketitle

The Ising universality class is one of the most robust paradigms of collective behavior. It describes the critical behavior of equilibrium systems exhibiting a $\mathbb Z_2$ symmetry, from uniaxial spins to liquid-gas phase separation \cite{fisher1967theory}. Ising-like phenomenology also naturally emerges in many living and non-equilibrium systems with bistable behavior, such as cell decision-making \cite{romeo2026information}, social behavior \cite{mullick2025sociophysics}, gene regulatory networks \cite{weber2016cellularising,simpson2023ising, sarra2025maximum}, neural populations \cite{hopfield1982neural, schneidman2006pairwise,  tkavcik2015thermodynamics, lynn2025exact, sarra2025maximum}, and cell membranes \cite{bagheri2026membrane}.
The equilibrium Ising universality class is remarkably robust to non-equilibrium perturbations \cite{bassler1994critical}, which manifest as forces or terms in the equations of motion that violate detailed balance \cite{tauber2014criticaldynamics}. Examples of systems that fall in the Ising class despite such violations include cellular automata \cite{grinstein1985pca}, kinetic spin systems coupled to thermal baths at different temperatures \cite{odor2004review}, or models with explicit nonreciprocal interactions \cite{dicarlo2025offequilibrium}. Similar conclusions hold for active systems undergoing motility-induced phase separation \cite{cates2015mips,caballero2018mips,maggi2022critical}.

The universality class is determined by the long-range behavior at the critical point, and breaking detailed balance at the microscopic scale does not guarantee a violation at large scales \cite{tauber2014criticaldynamics, tauber1997critical}. From a renormalization group perspective, this means that terms violating detailed balance often flow to zero at the RG fixed point \cite{bassler1994critical,akkineni2004nonequilibrium}. It might therefore be tempting to conclude that Ising criticality is robust to any non-equilibrium perturbation. Identifying mechanisms that overturn this conclusion is of great interest, as it may reveal new ingredients that dominate the large-scale behavior of non-equilibrium systems. This is especially relevant in the context of living systems, where collective behaviors are often tied to crucial biological functions \cite{downes1969swarming,sullivan1981insect,cresswell1994flocking,ladoux2017cells}.

At equilibrium, one known route away from Ising universality is disorder in the form of quenched impurities, which can alter the universality class in three dimensions \cite{harris1974random}. Likewise, spatiotemporal disorder is suspected to affect the critical behavior of a three-dimensional Ising model \cite{vojta2016spatiotemporal}. At equilibrium, however, such spatiotemporal impurities affect only the dynamic universality class, while leaving the static critical behavior unchanged \cite{hohenberg1977theory}. Hyperscaling relations between critical exponents are a fundamental consequence of equilibrium. In the equilibrium Ising class, correlation and response functions are related by the Fluctuation-Dissipation Theorem (FDT), which forces the exponents describing how the correlation ($\eta$) and the response function ($\eta^\prime=2-\gamma/\nu$) diverge to be the same. A splitting $\eta\neq\eta'$ therefore provides a definitive signature that non-equilibrium effects remain relevant even at the largest scales.

In this letter, we introduce the \textit{Brownian Ising model} (BIM), a minimal field theory of a non-equilibrium system in which an order parameter $\psi$ is coupled to an \emph{independently} evolving conserved density $\rho=\rho_{0}+\delta\rho$, which acts as strongly correlated spatiotemporal disorder. The absence of feedback from $\psi$ onto $\rho$ introduces a nonreciprocal structure that explicitly breaks detailed balance. We study the critical behavior of the BIM using the dynamical renormalization group and show that it \emph{does not} fall into the Ising universality class -- a complete derivation and further details are provided in the accompanying long paper~\cite{Scandolo2026BimLong}. 

The BIM describes a broad class of physical systems, including bistable collections of motile particles in the weak-advection regime -- such as the active Ising model \cite{solon2013flocking,solon2015flocking,chen2025bim,lima2022diffusivevoter} -- and spatially extended bistable reaction networks embedded in strongly fluctuating environments where motile catalysts locally modulate transition rates \cite{Scandolo2026BimLong}. The hydrodynamics couples standard Model A relaxation for $\psi$ to a diffusing conserved field $\rho$,
\begin{align}
    \lambda^{-1}\partial_t \psi &= \nabla^2\psi - r\psi - \tilde{u}\psi^3 + \sqrt{2\tilde{\lambda}\lambda^{-2}}\,\xi,
    \label{eq:originalModelA}\\
    D_0^{-1}\partial_t\delta\rho &= \nabla\cdot\!\left(-\nabla\delta\rho + \sqrt{2\tilde{D}_0 D_0^{-2}}\,\boldsymbol{\zeta}\right),
    \label{eq:rho}
\end{align}
where $\xi$ and $\boldsymbol{\zeta}$ are independent Gaussian white noises, and the coefficients of Eq.~\eqref{eq:originalModelA} are in principle all dependent on the local density $\rho$. Taylor expanding $\lambda(\rho)$, $r(\rho)$, $\tilde{u}(\rho)$, and $\tilde{\lambda}(\rho)$ around the mean density $\rho_0 = \tilde{D}_0/D_0$, it can be shown through a renormalization-group argument \cite{scandolo2023active} that the only relevant first-order term is $g_0\psi\,\delta\rho$, where $g_0 = r'(\rho_{0})$ is now independent on fluctuations of $\rho$. The equation of motion for $\psi$ retaining only RG-relevant terms is therefore
\begin{align}
    \lambda_0^{-1}\partial_t\psi &= \nabla^2\psi - r_0\psi - \tilde{u}_0\psi^3 - g_0\psi\,\delta\rho + \sqrt{\frac{2\tilde{\lambda}_0}{\lambda_0^{2}}}\,\xi,
    \label{eq:psi}
\end{align}
where $r_0$ is the (bare) distance to criticality, $\tilde{u}_0$ controls the strength of the non-linearity, $g_0$ is now an effective density-magnetization coupling, and the subscript zero denotes evaluation at the mean density $\rho_0$. Equations~\eqref{eq:rho}--\eqref{eq:psi} agree with exact hydrodynamics derived for active Ising models in the small-advection limit using the Doi-Peliti formalism \cite{kourbane2018exact,scandolo2023active, peliti1985path}. The conserved density evolves purely diffusively, with equal-time correlations $C_\rho(\mathbf{q},t) = \rho_0\,e^{-D_0 q^2 t}$ set by the mean density $\rho_0$, reflecting the Poissonian statistics of the underlying diffusion. The relevant coupling $g_0\psi\,\delta\rho$ enters through the mass term and effectively subjects the ordering dynamics to a spatio-temporally fluctuating distance to criticality -- structurally analogous to the field theory of the Ising model with quenched impurities \cite{khmelnitskii1975second,lubensky1975critical,grinstein1976disordered,grinstein1977dynamics}, but with a crucial distinction: the disorder here is \emph{dynamical} rather than static. This type of dynamical disorder is akin to the one studied in~\cite{vojta2016spatiotemporal}.

Both $\psi$ and $\delta\rho$ are hydrodynamic slow variables. Near criticality, $\psi$ undergoes critical slowing down with relaxation time $\tau_\psi \sim \lambda_0^{-1}\Lambda^{z-2}q^{-z}$, while density modes relax diffusively as $\tau_\rho \sim D_0^{-1}q^{-2}$. Their ratio,
\begin{equation}
    \frac{\tau_\rho}{\tau_\psi} = w_0\left(\frac{q}{\Lambda}\right)^{z-2}, \qquad w_0 \equiv \frac{\lambda_0}{D_0},
\end{equation}
controls the relative dynamics of the two fields and defines two physically distinct limits. When $w_0(q/\Lambda)^{z-2}\gg 1$, the density is effectively frozen on the timescales of $\psi$, yielding quenched disorder. The resulting theory maps onto Model A with a random quenched mass -- the universality class of the diluted Ising model \cite{lubensky1975critical,grinstein1976disordered, grinstein1977dynamics} -- but, as we show below, this limit is \emph{not} the fixed point of the BIM. In the opposite limit $w_0(q/\Lambda)^{z-2}\ll 1$, the density relaxes far faster than $\psi$. Over any intermediate timescale $\tau_\rho\ll\delta t\ll\tau_\psi$, density fluctuations fully decorrelate while $\psi$ barely evolves, so the order parameter experiences an effective noise with correlation \cite{Scandolo2026BimLong}
\begin{equation}
    C^\mathrm{eff}_\rho(\mathbf{q},t) = \frac{2\tilde{D}_0}{D_0^2\,q^2}\,\delta(t).
    \label{eq:effnoise}
\end{equation}
This noise is white in time \footnote{For $z>2$, the exponential kernel $e^{-D_{0}q^{2} t}$ naturally approaches $2D_{0}^{-1} q^{-2}\delta(t)$ as $q\ll w_{0}^{1/(2-z)}\Lambda$.} but carries long-range spatial correlations decaying as $C_{\rho}(r)\sim r^{-(d-2)}$. As we demonstrate below, this fast diffusion limit is recovered at the RG fixed point of the BIM. This follows from $z>2$, meaning that the strong timescale separation $w_0(q/\Lambda)^{z-2}\ll1$ naturally emerges in the hydrodynamic limit $q\to0$. Unlike short-range multiplicative noise, the long-range nature of $C^\mathrm{eff}_\rho$ is RG-relevant near $d_c=4$ and destabilizes the Ising fixed point, driving the system to a new universality class. The enhanced large-scale correlations induced by this long-range noise result in a negative anomalous dimension $\eta<0$: spatial correlations of $\psi$ decay more slowly than in the Gaussian theory, in sharp contrast to the standard Ising case where $\eta>0$.

The BIM is structurally distinct from Model C \cite{hohenberg1977theory}, which also couples a $\mathbb{Z}_2$ order parameter to a conserved scalar density. The key difference is that in the BIM the density current contains no magnetization-dependent term $\nabla\psi^2$, because agents move independently of their internal state. This asymmetry, given by the nonreciprocal structure of the interaction, is preserved under coarse-graining: no $\psi$-dependence of the density current can be generated by the RG if absent microscopically. Systems that do carry this feedback include ferromagnetic spin fluids with Lennard-Jones interactions \cite{nijmeijer1998isingfluid,nijmeijer1998heisenbergfluid} and active systems in which detailed balance is restored in the passive limit \cite{agranov2024thermodynamically}; these fall in the Model C universality class, where their weak violations of detailed balance are RG-irrelevant \cite{akkineni2004nonequilibrium}. The BIM therefore corresponds to a strongly detailed-balance-breaking limit of the general $\psi$-$\rho$ theory, for which the following analysis applies.

Analysis of the homogeneous solutions of Eqs.~\eqref{eq:rho}--\eqref{eq:psi} reveals a standard pitchfork bifurcation in $\psi_0$ as $r_0$ changes sign, with the disordered phase $\psi_0=0$ stable for $r_0>0$ and two ordered states $\psi_0=\pm\sqrt{-r_0/\tilde{u}_0}$ emerging for $r_0<0$. At mean-field level, $r_0$ directly measures the distance to criticality; fluctuations shift the critical point to $r_c<0$. We define the distance from the critical point $\tau_0 \equiv r_0 -r_c$. Unlike active Ising models \cite{solon2015flocking}, the homogeneous ordered state is stable with respect to spatially inhomogeneous fluctuations, confirming that the transition is continuous.

We analyze the critical dynamics of Eqs.~\eqref{eq:psi}--\eqref{eq:rho} using the Martin-Siggia-Rose path-integral formalism \cite{martin1973statistical, tauber2014criticaldynamics}, which introduces auxiliary response fields $\tilde{\psi}$ and $\tilde{\rho}$. The field-theoretical RG \cite{martin1973statistical,de1976techniques,janssen1976on, hohenberg1977theory} near $d_c=4$ then requires defining $Z$ factors to absorb UV divergences and relate bare to renormalized quantities; their logarithmic derivatives with respect to the RG scale $\mu$ define the anomalous dimensions $\gamma_a \equiv \mu\partial_\mu\ln Z_a$, from which the beta functions $\beta_a \equiv \mu\partial_\mu a_{R}$ and critical exponents follow\cite{peskin1995introduction, tauber2014criticaldynamics, Scandolo2026BimLong}. The three dimensionless coupling constants of the theory are
\begin{equation}
    u_R = Z_u \frac{\tilde{\lambda}_0}{\lambda_0}\,\tilde{u}_0\,\mu^{-\epsilon}, \quad f_R = Z_f \frac{\tilde{D}_0}{D_0}\,g_0^2\,\mu^{-\epsilon}, \quad w_R =Z_w \frac{\lambda_0}{D_0},
\end{equation}
where $\mu$ is the RG momentum scale and $\epsilon=4-d$. The parameter $u_R$ controls the $\psi^4$ non-linearity, $f_R$ encodes the density-magnetization coupling, and $w_R$ is the timescale ratio introduced above. Similarly, we define the renormalized distance from the critical point, $\tau_R = Z_\tau \tau_0 \mu^{-2}$. 
In systems with multiple dynamical timescales -- here $\tau_\rho$ and $\tau_\psi$ -- the natural dimensionless coupling that flows to a nontrivial fixed point can differ between regimes \cite{FREY199552, frey1990renormalized, cavagna2019renormalization, cavagna2019dynamical}. We therefore define
\begin{equation}
 X_R = \frac{w_R}{1+w_R} \qquad , \qquad \tilde{f}_R = X_R f_R,
\label{eq:Xtildef}
\end{equation}
which ensures that $\tilde{f}_R$ remains finite in both the fast-diffusion limit $w_R\to 0$ and the quenched limit $w_R\to\infty$. A fixed point at intermediate values $0 < X_R^\star < 1$  -- corresponding to finite $w_R^\star$ -- would require $\gamma_\lambda^\star = 0$, i.e., strong dynamic scaling\cite{hohenberg1977theory, tauber2014criticaldynamics} with $z=2$. At one-loop the beta functions for $u_{R},\tilde f_{R},X_{R}$ are,
\begin{align}
    \beta_{\tilde{f}} &= \tilde f_{R} \left(\epsilon+ \tilde f_{R} (X_{R}+1) X_{R}^{2}-6 u_{R} \right) \label{eq:betaf}\\
   %\beta_{\tilde{f}} &= \tilde{f}_R\!\left(\epsilon - 6u_R + \frac{w_R^2}{(1+w_R)^3}\tilde{f}_R\right), \\
    \beta_{u} &= u_{R} \!\left(\epsilon -9 u_{R}+2 \tilde f_{R} \left(X_{R}^2+2\right)\right) \label{eq:betau}\\
    %\beta_{u} &= u_R\!\left(\epsilon - 9u_R + \frac{4+8w_R+6w_R^2}{(1+w_R)^3}\tilde{f}_R\right), \\
    \beta_{X} &= -\tilde f_{R} X_{R}^{3} (1-X_{R}) \label{eq:betaw}
    %\beta_{w} &= -\frac{w_R^3}{(1+w_R)^3}\,\tilde{f}_R. \label{eq:betaw}
\end{align}
\begin{figure}[t]
    \includegraphics[width=\columnwidth]{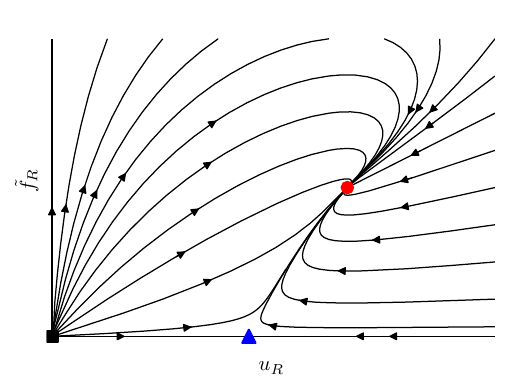}
    \caption{{\bf One-loop renormalization-group flow} in the $(u_R,\tilde{f}_R)$ couplings plane at $X_R = 0$ . The black square marks the Gaussian fixed point, the blue triangle the Wilson-Fisher (Model A) fixed point, and the red circle the BIM fixed point. The Wilson-Fisher fixed point is unstable with respect to the off-equilibrium perturbation. }
    \label{fig:Rgflow}
\end{figure}
The Model A fixed point ($\tilde{f}_R^{\star}=0$, $u_R^{\star}=\epsilon/9$) is IR-unstable to perturbations in $\tilde{f}_R$: any nonzero density coupling drives the system away from Ising universality, as shown in Fig.~\ref{fig:Rgflow}. A novel IR-stable fixed point emerges,
\begin{equation}
    X_R^{\star} = 0, \qquad u_R^{\star} = \frac{\epsilon}{6}, \qquad \tilde{f}_R^{\star} = \frac{\epsilon}{8},
    \label{eq:BIMfp}
\end{equation}
which we refer to as the BIM fixed point. The critical exponents follow from the anomalous dimensions $\gamma_a^{\star}$ evaluated at this fixed point via $\nu=(2-\gamma_\tau^{\star})^{-1}$, $\eta=-\gamma_\psi^{\star}$, $\eta'\equiv-(\gamma_\psi^{\star}+\gamma_{\tilde\psi}^{\star})/2$ and $z=2+\gamma_\lambda^{\star}$:
\begin{align}
    \nu &= \tfrac{1}{2}+\tfrac{\epsilon}{8}+\mathcal{O}(\epsilon^2), & z &= 2+\mathcal{O}(\epsilon^2),
    \label{eq:BIMFP1}
    \\
    \eta &= -\tfrac{\epsilon}{8}+\mathcal{O}(\epsilon^2), & \eta' &= \mathcal{O}(\epsilon^2).
    \label{eq:BIMFP2}
\end{align}
These results deviate from the Ising class in two fundamental ways. First, the correlation-length exponent $\nu\neq\frac{1}{2}+\frac{\epsilon}{12}$ already at one loop. Second, and more fundamentally, $\eta\neq\eta'$: at any equilibrium fixed point these two exponents are equal because the FDT equates correlation and response functions. Their splitting here is a direct quantitative signature that the BIM fixed point is genuinely non-equilibrium. The negative sign of $\eta$ reflects the long-range nature of $C_\rho^\mathrm{eff}$ [Eq.~\eqref{eq:effnoise}]: at $X_R^{\star}=0$ ($w_R^{\star}=0$) the density coupling reduces to a spatially long-range multiplicative noise that enhances correlations beyond the Gaussian level.

\begin{table*}[t]
\centering
\begin{tabular*}{\textwidth}{@{\extracolsep{\fill}}r|c c c c c}
\hline
\hline
FP & $\nu$ & $\eta$ & $\eta'$ & $z$ & $\omega$ \\[4pt]
\hline
Gaussian & $\frac{1}{2}$ & $0$ & $0$ & $2$ & $0$ \\[4pt]
\hline
Model A & $\frac{1}{2}+\frac{\epsilon}{12}$ & $\frac{\epsilon^2}{54}$ & $\frac{\epsilon^2}{54}$ & $2+\frac{\epsilon^2}{54}\!\left(6\log\frac{4}{3}-1\right)$ & $\epsilon$ \\[4pt]
\hline
BIM & $\frac12+\frac\epsilon8+\frac{\epsilon ^2}{768}  \left(29-24 \log \left(\frac{4}{3}\right)\right)$ & $- \frac \epsilon 8$ 
 & $\frac{\epsilon^2}{96}\!\left(5-12\log\frac{4}{3}\right)$ & $2+\frac{\epsilon^2}{96}\!\left(24\log\frac{4}{3}-5\right)$ & $\frac{\epsilon^2}{96}\!\left(24\log\frac{4}{3}-5\right)$ \\[4pt]
\hline
\end{tabular*}
\caption{{\bf Critical exponents} at the Gaussian, Model A, and BIM fixed points near $d_c=4$, to lowest nontrivial order in $\epsilon=4-d$. The exponent $\omega$ is the correction-to-scaling exponent within the attractive manifold. Note that $\eta\neq\eta'$ at the BIM fixed point, signaling a violation of the fluctuation-dissipation theorem and a genuinely non-equilibrium universality class.
}
\label{tab:UC}
\end{table*}

The stability of the BIM fixed point is not fully resolved at one loop. Near $X_R^{\star}=0$, Eq.~\eqref{eq:betaw} gives $\beta_X\sim -X_R^3\tilde{f}_R^{\star}$, so the linearized eigenvalue in the $X_R$ direction vanishes: $X_R$ approaches its fixed-point value polynomially in RG time rather than exponentially. This implies that the sign of the correction-to-scaling exponent $\omega=\mathcal{O}(\epsilon^2)$, which determines whether the fixed point is stable ($\omega>0$) or unstable ($\omega<0$), cannot be resolved at one loop. Stability is confirmed at two loops: causality restricts the relevant two-loop diagrams to just two at $w_R^{\star}=0$, yielding\cite{Scandolo2026BimLong}
\begin{equation}
    z = 2+\frac{\epsilon^2}{96}\!\left(24\log\tfrac{4}{3}-5\right), \qquad \omega = \frac{\epsilon^2}{96}\!\left(24\log\tfrac{4}{3}-5\right).
\end{equation}
Since $24\log(4/3)>5$, we conclude $\gamma_\lambda^{\star}>0$ and hence the BIM fixed point is stable near $d=4$. The exponent $\omega$ determines how quickly the RG flow converges to the fixed point, therefore a small value $\omega\approx0.020$ at $d=3$ indicates relatively slow convergence to asymptotic scaling. A full summary of exponents at all accessible fixed points to lowest nontrivial order in $\epsilon$ is given in Table~\ref{tab:UC}.

We now discuss several exact results, valid to all orders in perturbation theory, which establish that both the Ising and diluted-Ising fixed points are unstable for $d=3$, making the BIM fixed point the only viable candidate to describe the large-scale critical behavior of Eqs.~\eqref{eq:originalModelA}--\eqref{eq:rho}. Two structural properties of the field theory are key. First, the linearity of Eq.~\eqref{eq:rho} implies that the density sector does not renormalize and leads to the following relation \cite{Scandolo2026BimLong},  %independently giving $Z_f=Z_\rho$, and, crucially,
\begin{equation}
    Z_w = Z_\lambda \implies \beta_X = -X_R(1-X_{R})\,\gamma_\lambda.
    \label{eq:noDren}
\end{equation}
Second, the field theory in Eq.~\eqref{eq:rho}-\eqref{eq:psi} possesses a shift symmetry: $\rho\to\rho+c$, $\tau\to\tau-c$ leaves the equations invariant for any constant $c$. This forces $Z_f=Z_\tau^2$ \cite{Scandolo2026BimLong} and therefore
\begin{equation}
    \beta_{\tilde f}=\tilde f_R (\epsilon-2\gamma_{\tau}+(X_R-1)\gamma_{\lambda})
    \label{eq:betafexact}
\end{equation}
Both relations hold to all orders in perturbation theory.

\textit{Instability of the diluted-Ising fixed point.} Equation~\eqref{eq:noDren} gives $\beta_X^{\star}=X_R^{\star}(1-X_R^{\star})(2-z)$ at any fixed point, so a stable fixed point at $X^{\star}_R=1$ (quenched disorder) requires $z<2$. The diluted Ising model is known to have $z>2$ both near $d=4$ \cite{grinstein1977dynamics} and in $d=3$ \cite{hasenbusch2007relaxational}, ruling out $X^{\star}_R=1$ as a stable fixed point.

\textit{Instability of the Ising fixed point.} At the Ising fixed point ($\tilde f_R^{\star}=0$), Eq.~\eqref{eq:betafexact} implies that stability requires 
\begin{equation}
    \begin{split}
        \epsilon &< 2\gamma_\tau^{\star} - (X_R^{\star}-1)\gamma_{\lambda}^{\star} =\\
        &=2(2-\nu^{-1})-(X_R^{\star}-1)(z-2)
    \end{split}
\end{equation}
i.e., 
\begin{equation}
    (d+(1-X_R^{\star})(z-2)) \nu >2 \ .
\end{equation}
For $X_{R}^{\star}>0$, $\beta_{X}^{\star}=0$ requires $(1-X_R^{\star})(z-2)=0$. This recovers the Harris criterion of stability $d\nu>2$ in the presence of quenched disorder ($X_R^\star=1$) \cite{harris1974random}. If, on the other hand $X_R^{\star}=0$, as happens for Model A (since $z>2$), the $\tilde f_R^{\star}=0$ is stable only if
\begin{equation}
    (d+z-2)\nu>2
\end{equation}
in agreement with the modified Harris criterion proposed in \cite{vojta2016spatiotemporal}. Using $\nu\approx0.630$ and $z\approx2.02$ for Model A in $d=3$ \cite{pelissetto2002critical,hasenbusch2020dynamic,adzhemyan2022dynamic}, we conclude that $\tilde f_{R}^{\star}=0$ is \emph{unstable} \footnote{In principle, nothing prevents a fixed point with $0<X_R^{\star}<1$ and $z=2$. However, no such fixed points have been found within the present perturbative analysis. Were it to exist, a fixed point with $z=2$ would still obey the same stability criterion $d\nu>2$, despite the disorder not being quenched ($X_R^\star\neq1$). This agrees with the findings of \cite{vojta2016spatiotemporal}. A fantasy fixed point with Ising static exponents ($\tilde f_{R}^{\star}=0$) and $z=2$ would thus also be unstable all the way down to $d=3$.}. The same argument predicts that the Ising fixed point is stable for $d=2$, in agreement with previous numerical results \cite{solon2015flocking,chen2025bim}.

\begin{figure}[t]
    \includegraphics[width=\columnwidth]{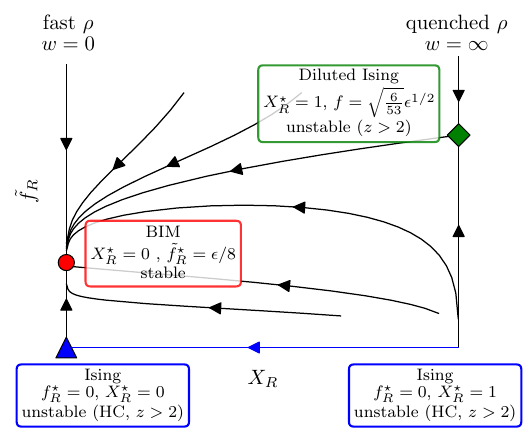}
    \caption{{\bf Qualitative sketch of the full renormalization group flow} The Ising fixed point (defined by the blue line $f=0$, $\forall w$) is globally unstable across all regimes. For $w=\infty$, the equilibrium fixed point is unstable according to the Harris criterion. For any finite $w$, the Ising point remains unstable because its dynamical exponent satisfies $z>2$. In the $w=0$ limit, instability is instead driven by the modified Harris criterion. Furthermore, the diluted Ising fixed point is unstable because $z>2$. Since these results are non-perturbative, the RG flow must terminate at a different attractor. The BIM fixed point emerges as the only viable candidate to govern the critical dynamics, as its dynamical exponent is both physically compatible and satisfies the modified Harris stability criterion.}
    \label{fig:mock}
\end{figure}

\textit{Exact scaling relation at the BIM fixed point.} Since neither of the diluted or standard Ising fixed points is stable, the BIM fixed point must govern the critical behavior for $d\leq4$. A stable fixed point with finite $\tilde{f}_R^{\star}$ requires $\beta_{\tilde{f}}^{\star}=0$, implying that at any $\tilde f_R^{\star}\neq0$ fixed point the following exact relation holds,
\begin{equation}
    \nu = \frac{2}{d+z-2}
    \label{eq:exactnu}
\end{equation}
 Combined with the two-loop result for $z$, this fixes the $\epsilon^2$ correction to $\nu$ as well, giving in $d=3$:
\begin{equation}
    \nu = \frac12+\frac\epsilon8+\frac{\epsilon ^2}{768}  \left(29-24 \log \left(\frac{4}{3}\right)\right)\approx 0.65
\end{equation}

Figure~\ref{fig:mock} summarizes the non-perturbative structure of the RG flow in the $(X_R,\tilde{f}_R)$ plane, inferred from the stability criteria derived above. The flow is anchored at $X_R=0$ and $X_R=1$ by the one-loop beta functions; the interpolating trajectories are schematic. As we have previously discussed, the stability of the BIM fixed point at $X_R^\star=0$ is not resolved at one loop -- where $\beta_X \sim X_R^3$ vanishes polynomially -- but is confirmed by a two-loop calculation \cite{Scandolo2026BimLong}. The crossover between the two fixed points cannot be captured quantitatively within the standard Callan-Symanzik framework, and requires a double expansion in both $\epsilon$ and $1/w_R$, which is beyond the scope of this work \cite{FREY199552, frey1990renormalized, cavagna2019renormalization}.

\textit{Slow convergence and finite-size corrections.} The small value of the correction-to-scaling exponent, $\omega\approx0.020$ at $d=3$, has an important practical consequence. Finite-size corrections to scaling observables decay as $L^{-\omega}$; with $\omega$ this small, these corrections remain substantial even for very large system sizes. Numerical studies of BIM criticality should therefore anticipate very strong deviations from asymptotic scaling and will require either exceptionally large systems or explicit correction-to-scaling analyses to reliably extract the critical exponents.

{\it Conclusions} We have shown that coupling an Ising order parameter to a conserved diffusive field-without feedback from the order parameter onto the density-constitutes a qualitatively new route out of Ising universality. The mechanism is identified by the renormalization group: in the fast-diffusion limit, which is the IR fixed point ($w_R^{\star}=0$), the density acts as an effective multiplicative noise with long-range spatial correlations $C_\rho^\mathrm{eff}\sim q^{-2}\delta(t)$. Unlike short-range multiplicative noise, this long-range noise is RG-relevant near $d_c=4$ and drives the system to a novel fixed point-the BIM universality class-with $u_R^{\star}=\epsilon/6$, $\tilde{f}_R^{\star}=\epsilon/8$.
The BIM universality class is distinguished from the Ising class by two concrete signatures. First, the correlation-length exponent $\nu=\frac{1}{2}+\frac{\epsilon}{8}$ deviates from the Ising value $\frac{1}{2}+\frac{\epsilon}{12}$ already at one loop, yielding at two-loops order $\nu\approx0.65$ in $d=3$. Second, and more fundamentally, the anomalous dimensions of the correlation and response functions are distinct, $\eta\neq\eta'$, in direct violation of the fluctuation-dissipation theorem: $\eta=-\epsilon/8$ is negative and already one-loop, while $\eta'=\mathcal{O}(\epsilon^2)>0$. This splitting is both the defining hallmark of the new universality class and a directly measurable quantity in simulations of microscopic realizations. Going beyond perturbation theory, a generalized Harris stability criterion $(d+z-2)\nu>2$, valid to all orders in perturbation theory, provides a further consistency check. We establish that neither the Ising nor the diluted-Ising fixed points are viable in $d=3$, while the same Harris-like analysis predicts that the Ising universality class is stable in $d=2$-in agreement with existing numerical results \cite{solon2015flocking,chen2025bim}.
The correction-to-scaling exponent $\omega\approx0.020$ in $d=3$ is anomalously small, implying that finite-size corrections, which decay as $L^{-\omega}$ with the system size $L$, will persist even for very large systems. Numerical studies of BIM criticality should anticipate this and incorporate explicit correction-to-scaling analyses. More broadly, the BIM mechanism -- $\mathbb{Z}_2$ symmetry combined with nonreciprocal coupling to a conserved density -- is likely to be realized in a wide range of motile, driven, and environmentally fluctuating systems, from cell populations and neural assemblies to bird flocks and reaction networks with mobile catalysts. In all such cases, the FDT splitting $\eta\neq\eta'$ is the cleanest observable signature that the system has escaped Ising universality.

\section{Acknowledgment}
We thank Rosalba Garcia-Millan for discussions during the early stages of this work. MS and LDC thank Andrea Cavagna, Tomas S. Grigera for useful discussion. LDC thanks Q. Yu for useful discussions. MS is grateful to the Department of Applied Mathematics and Theoretical Physics at the University of Cambridge, where the early stage of this work was performed, for their support. MS acknowledges support from ERC Advanced Grant RG.BIO (no. 785932). LDC was supported by the National Science Foundation, through the Center for the Physics of Biological Function (PHY--1734030).

\bibliographystyle{apsrev4-2}
\bibliography{references}

\end{document}